\documentclass[sigconf]{acmart}
\usepackage{algorithm}
\usepackage{algpseudocode}
\usepackage{multirow}
\usepackage{subcaption}
\usepackage[normalem]{ulem}
\usepackage[skip=10pt,font=large]{caption}
\usepackage{enumitem}
\usepackage{tikz}
\setlist[itemize]{leftmargin=*}

\AtBeginDocument{%
  \providecommand\BibTeX{{%
    \normalfont B\kern-0.5em{\scshape i\kern-0.25em b}\kern-0.8em\TeX}}}

\copyrightyear{2023} 
\acmYear{2023} 
\setcopyright{acmlicensed}\acmConference[SIGIR-AP '23]{Annual International ACM SIGIR Conference on Research and Development in Information Retrieval in the Asia Pacific Region}{November 26--28, 2023}{Beijing, China}
\acmBooktitle{Annual International ACM SIGIR Conference on Research and Development in Information Retrieval in the Asia Pacific Region (SIGIR-AP '23), November 26--28, 2023, Beijing, China}
\acmPrice{15.00}
\acmDOI{10.1145/3624918.3625339}
\acmISBN{979-8-4007-0408-6/23/11}

\begin{document}

\title{How to Index Item IDs for Recommendation Foundation Models}

\settopmatter{authorsperrow=4}

\author{Wenyue Hua}
\affiliation{%
  \institution{Rutgers University}
  \country{}
}
\email{wenyue.hua@rutgers.edu}

\author{Shuyuan Xu}
\affiliation{%
  \institution{Rutgers University}
  \country{}
}
\email{shuyuan.xu@rutgers.edu}

\author{Yingqiang Ge}
\affiliation{%
  \institution{Rutgers University}
  \country{}
}
\email{yingqiang.ge@rutgers.edu}

\author{Yongfeng Zhang}
\affiliation{%
  \institution{Rutgers University}
  \country{}
}
\email{yongfeng.zhang@rutgers.edu}

\renewcommand{\shortauthors}{Wenyue Hua, Shuyuan Xu, Yingqiang Ge, and Yongfeng Zhang}

\begin{abstract}
Recommendation foundation model utilizes large language models (LLM) for recommendation by converting recommendation tasks into natural language tasks. It enables generative recommendation which directly generates the item(s) to recommend rather than calculating a ranking score for each and every candidate item as in traditional recommendation models, simplifying the recommendation pipeline from multi-stage filtering to single-stage filtering. To avoid generating excessively long text and hallucinated recommendations when deciding which item(s) to recommend, creating LLM-compatible item IDs to uniquely identify each item is essential for recommendation foundation models. In this study, we systematically examine the item ID creation and indexing problem for recommendation foundation models, using P5 as an example of the backbone LLM. To emphasize the importance of item indexing, we first discuss the issues of several trivial item indexing methods, such as random indexing, title indexing, and independent indexing. We then propose four simple yet effective solutions, including sequential indexing, collaborative indexing, semantic (content-based) indexing, and hybrid indexing. Our study highlights the significant influence of item indexing methods on the performance of LLM-based recommendation, and our results on real-world datasets validate the effectiveness of our proposed solutions. The research also demonstrates how recent advances on language modeling and traditional IR principles such as indexing can help each other for better learning and inference. Source code and data are available at \url{https://github.com/Wenyueh/LLM-RecSys-ID}.
\end{abstract}



\maketitle

\section{Introduction}
Foundation Models such as Large Language Models (LLMs) \cite{chung2022scaling,chowdhery2022palm,touvron2023llama} have significantly impacted research areas such as natural language processing (NLP) and computer vision (CV) \cite{lin2022learning}, and have been applied to various recommender system (RS) tasks. Recent research such as P5 \cite{p5} and M6Rec \cite{m6} leverage the advantages of pre-trained LLMs for recommendation \cite{li2023large}: they incorporate rich user behavior and knowledge information into pre-training and benefit from the strong learning ability of foundation models for recommendation. Pre-trained LLMs also have improved reasoning ability \cite{huang2022towards} to infer user interests based on the context. Therefore, these models aim to utilize LLMs pre-trained on extensive natural language corpora for RS by transforming recommendation tasks into language generation tasks, enabling generative recommendation.

Since item description may include a large number of words (e.g., a product title/description could include tens/hundreds of words and a news article could include thousands of words), we can hardly expect an LLM to generate the complete and exact item description when deciding which item(s) to recommend, because the generated text may not even correspond to a real existing item in the item database, leading to the hallucination problem \cite{lin2021truthfulqa, evans2021truthful} in LLM-based recommendation. As a result, it is important to assign a unique ID to each item so that each item is represented by a small number of characteristic tokens while being distinguishable from each other. For example, a business location in Yelp may be assigned the ID ``location\_4332'' and be further represented as a sequence of tokens such as $\langle$location$\rangle$$\langle$\_$\rangle$$\langle$43$\rangle$$\langle$32$\rangle$ \cite{p5}. Note that the item ID may not necessarily be number tokens, rather, as long as it is a unique identifier for an item, then it may be considered as an ID for the item. For example, the title of the movie ``The Lord of the Rings'' can be considered as the ID of the movie, which consists of a sequence of word tokens rather than number tokens. The ID may even be a sequence of words that do not convey an explicit meaning, e.g., ``ring epic journey fellowship adventure''. 

However, assigning LLM-compatible IDs to items is not a trivial task. First, there could be a huge amount of or even infinite items while each item should be assigned a unique ID so that items are distinguishable from each other for the foundation model. Second, item IDs should be compatible with natural language so that IDs can be integrated into natural language instructions for the pre-training, fine-tuning and prompting of LLMs. Third, trivial item indexing methods such as random indexing may not help and may even hurt the recommendation foundation models since they may mistakenly assign related IDs to unrelated items, misleading the training and prompting of LLMs. 
As a result, a comprehensive examination for LLM-oriented item indexing is needed, which enables the seamless adaptation of recommendation tasks to be compatible with LLMs, harnessing the potential of LLMs for recommendation.

Besides, a natural idea to ensure the generated text align with real items so as to avoid the hallucination problem is to employ a constrained decoding method \cite{autoregressive}. However, utilizing constrained generation for free-form long text is impractical. This is because constrained decoding essentially prescribes a singular mode of expressing the content, negating the flexible nature of long-text narratives. By compelling the model to adhere to specific descriptive patterns, the model is required to memorize rigid text patterns in addition to recommendation-specific knowledge. This additional complexity can dilute the primary purpose of the model and hinder its efficacy in performing the core recommendation task.

Motivated by the above reasons, this paper concentrates on the item indexing problem for LLM-based recommenders: how to assign a unique ID (i.e., token sequence) for each item. We study the issue based on P5 \cite{p5}, a representative LLM for RS model. P5 employs pre-training over foundation models and converts recommendation tasks into natural language sentences based on personalized prompts. We first experiment on three trivial indexing methods and show their limitations, some of which were employed in previous models: Independent Indexing (IID), Title Indexing (TID), and Random Indexing (RID). Based on the analysis, we further explore four novel indexing techniques: Sequential Indexing (SID), Collaborative Indexing (CID), Semantic (content-based) Indexing (SemID), and Hybrid Indexing (HID). To ensure the generated IDs align with real items during the recommendation stage so as to avoid hallucination, we develop a constrained decoding method \cite{autoregressive}, which is facilitated by crafting a prefix tree (a.k.a a trie) from the set of valid IDs and setting the generation probability of non-existent IDs to zero during the decoding phase. We show the performance of various ID methods on three widely-used datasets (Amazon Beauty, Amazon Sports, Yelp) and provide insights about the performance of different methods for LLM-based recommendation models.


\section{Related Work}

Many traditional recommendation models use a matching-based paradigm \cite{jannach2010recommender,zhang2019deep,adomavicius2005toward,koren2009matrix,zhang2017joint,xue2017deep,chen2021neural}. 
They project users (or user behavior history) and items into a shared embedding space 
and then estimate a user's preference for an item by calculating the ranking score using their embedding vectors, such as the inner product between the user and item vectors in matrix factorization \cite{koren2009matrix}.
Usually, this involves calculating ranking scores for each and every candidate item, making the matching and sorting process time consuming when the item pool is large \cite{zhao2020revisiting}. As a result, industrial RS usually has to use the multi-stage (usually two-stage) filtering pipeline \cite{covington2016deep}, where simple and efficient filtering methods such as rule-based filtering methods are used at early stages, while advanced filtering methods are used at later stages where candidate items are fewer. As a result, the most advanced models are only applied on a small subset of items.

Recently, there have been multiple attempts to pre-train foundational models for generative recommendation, which spare the expensive one-by-one candidate item matching process and instead directly generate the item to recommend. For example, P5 \cite{p5} unifies diverse recommendation tasks as natural language generation tasks within a sequence-to-sequence generation framework. Recommendation data such as user-item interactions, user descriptions, item metadata, and user reviews are converted to a common format---natural language sequences---using multiple personalized prompt templates. 
Each user or item is represented by a unique sequence of tokens as the user or item ID.
M6Rec \cite{m6} converts various recommendation tasks, such as content supply, delivery, and presentation, into natural language understanding or generation tasks. Input prompts incorporate user attributes, past behaviors, and detailed item descriptions provided by sellers. 
Users and items are represented as pre-computed embeddings from their attributes and descriptions. 
LMRecSys \cite{haowang} converts item-based recommendation tasks to text-based cloze tasks. The model is tested on the MovieLens-1M dataset
\cite{harper2015movielens}, which includes movies that pre-trained LLMs may have seen in web text. Items are represented by their titles that function as indices. 
This indexing method negatively affects the model performance as reported in the original paper: LLMs are not only ineffective for inferring the probability distribution of a multi-token span, but also the linguistic bias contained in titles may mislead the model as the title could contain little information about the content of the movie.

The three models use different methods to index items: P5 uses number tokens, M6Rec uses metadata-based embeddings, and LMRecSys uses item titles. This paper studies different item indexing methods under the LLM-based generative recommendation framework using P5 as an example backbone, which compares the effectiveness of different indexing methods, sheds light on the relationship between item indexing and foundation model pre-training, and also provides insights about which item indexing methods are most suitable for pre-training recommendation foundation models.

\vspace{-1ex}
\section{Preliminaries and Preceding Study}
\subsection{Introduction of P5 Paradigm}
This paper studies the indexing problem based on P5 \cite{p5}. P5 is a representative recommendation foundation model which enhances the generalization capabilities of existing recommender systems by integrating various tasks and personalized instruction prompts to pre-train a foundation model for recommendation. These tasks include sequential recommendation, rating prediction, explanation generation, review summarization, and direct recommendation. P5 is trained using input-target pairs of texts generated from a collection of prompt templates featuring personalized fields for distinct users and items: an example input prompt for sequential recommendation can be a description of user-item interactions such as ``According to the places user\_1 has visited: location\_1123, location\_4332, location\_8463, location\_12312, can you recommend another place for the user?'' and the output text is the next generated item such as ``Output: location\_1934''. In this study, we focus on the sequential recommendation task since it explicitly relies on the item interactions presented in the input prompt, making it highly sensitive to different indexing methods. 


\vspace{-1ex}
\subsection{The Angle Bracket Notation}

In this paper, we need to introduce Out-of-vocabulary (OOV) tokens to construct item indices in some indexing methods, 
which are tokens that are not part of the normal vocabulary of the language model. In our case, they are tokens that do not exist in the default T5 vocabulary \cite{raffel2020exploring}. To distinguish the newly created OOV tokens from existent tokens, we use angle brackets ``$\langle\rangle$'' to represent the newly created OOV token, and use text without ``$\langle\rangle$'' to represent an existent token in the default tokenizer. All OOV tokens are randomly initialized in the model and thus the text enclosed in ``$\langle\rangle$'' does not influence embeddings of the OOV tokens. The text within angle brackets ``$\langle\rangle$'' could be words or numbers, but no matter which case, the text within angle brackets only functions to distinguish different OOV tokens and it is irrelevant to the existent tokens. For example, $\langle$restaurant$\rangle$ $\langle$Greek$\rangle$ $\langle$2$\rangle$ is the index for an item in Yelp consisting three OOV tokens, where $\langle$restaurant$\rangle$ is a different token from the plain English word ``restaurant'', and $\langle$2$\rangle$ is a different token from the number ``2''. When we need to use the existent plain word tokens, we will use them without the angle brackets, such as ``restaurant'' and ``2''.

\subsection{Data Format and Prepossessing}
\label{sec:dataset}
Experiments are conducted on Amazon Sports $\&$ Outdoors, Amazon Beauty, and the Yelp dataset. The Amazon datasets \cite{ni2019justifying}\footnote{https://cseweb.ucsd.edu/~jmcauley/datasets/amazon\_v2/} are sourced from Amazon.com for product recommendations, while the Yelp dataset\footnote{https://www.yelp.com/dataset} provides a collection of user ratings and reviews for business recommendation. We use transaction records from Jan 1, 2019 to Dec 31, 2019, as in the original P5 paper \cite{p5}. The detailed statistics for these datasets can be found in Table \ref{tab:datasets}.

\begin{table}[t]
\vspace{-10pt}
    \centering
    \begin{tabular}{lcccc}
    \toprule
     & $\#$User & $\#$Item & $\#$Interactions & Sparsity($\%$) \\
     \hline
    Sports & 35,598 & 18,357 & 296,337 & 0.0453\\
    \hline
    Beauty & 22,363 & 12,101 & 198,502 & 0.0734 \\
    \hline 
    Yelp & 30,431 & 20,033 & 316,354 & 0.0519\\
    \bottomrule
\end{tabular}
\caption{Basic statistics of datasets}
\label{tab:datasets}
\vspace{-20pt}
\end{table}

\begin{table*}[t]
    \centering
    \setlength{\tabcolsep}{1.5pt}
    \begin{tabular}{lcccccccccccc}
    \toprule
        \multirow{2}{*}{Method} & \multicolumn{4}{c}{\textbf{Amazon Sports}} & \multicolumn{4}{c}{\textbf{Amazon Beauty}} & \multicolumn{4}{c}{\textbf{Yelp}}\\
        \cmidrule(lr){2-5} \cmidrule(lr){6-9} \cmidrule(lr){10-13}
         & HR@5 & NCDG@5 & HR@10 & NCDG@10 & HR@5 & NCDG@5 & HR@10 & NCDG@10 & HR@5 & NCDG@5 & HR@10 & NCDG@10 \\
        \midrule
        SASRec & 0.0233 & \uwave{0.0154} & 0.0350 & 0.0192 & 0.0387 & \uwave{0.0249} & 0.0605 & 0.0318 & 0.0170 & 0.0110 & 0.0284 & 0.0147 \\
        S$^3$\text{-Rec} & \uwave{0.0251} & \textbf{0.0161} & \uwave{0.0385} & \textbf{0.0204} & \uwave{0.0387} & 0.0244 & \textbf{0.0647} & \uwave{0.0327} & 0.0201 & 0.0123 & \uwave{0.0341} & 0.0168 \\
        \midrule
        RID & 0.0208 & 0.0122 & 0.0288 & 0.0153 & 0.0213 & 0.0178 & 0.0479 & 0.0277 & \uwave{0.0225} & \textbf{0.0159} & 0.0329 & \uwave{0.0193}\\
        TID & 0.0000 & 0.0000 & 0.0000 & 0.0000 & 0.0182 & 0.0132 & 0.0432 & 0.0254 & 0.0058 & 0.0040 & 0.0086 & 0.0049\\
        IID & \textbf{0.0268} & 0.0151 & \textbf{0.0386} & \uwave{0.0195} & \textbf{0.0394} & \textbf{0.0268} & \uwave{0.0615} & \textbf{0.0341} & \textbf{0.0232} & \uwave{0.0146} & \textbf{0.0393} & \textbf{0.0197}\\
        \bottomrule
    \end{tabular}
    \caption{Performances of the trivial indexing methods for P5 as well as the baselines. The numbers in bold represent the best results, while the numbers with a wave represent the second-best results. The results for RID and TID are significantly worse on Sports and Beauty, with a $p$-value < 0.05 under the paired Student's t-test protocol.}
    \label{tab:basic}
    \vspace{-15pt}
\end{table*}

These datasets organize user-item interactions by individual users. We split the datasets into training, validation, and testing by the frequently used leave-one-out setting: for each user's interaction sequence, we put the second-to-last item into the validation set, put the last item into the testing set, and put all other items of the sequence into the training set. For example, suppose the interaction sequence of user$_i$ is \{item$_{i,1}$, item$_{i,2}$, item$_{i,3}$, $\cdots$, item$_{i,k-1}$, item$_{i,k}$\}. Then the prediction of item$_{i,k-1}$ based on the sequence \{item$_{i,1}$, item$_{i,2}$, item$_{i,3}$, $\cdots$, item$_{i,k-2}$\} is used for validation and the prediction of item$_{i,k}$ based on the sequence \{item$_{i,1}$, item$_{i,2}$, item$_{i,3}$, $\cdots$, item$_{i,k-1}$\} is used for testing.

\subsection{Motivating Analysis of Item Indexing}
We motivate the exploration of indexing methods starting from three trivial indexing methods:
\begin{itemize}
    \item Random Indexing (RID): Assigning each item with a random number as the item ID. The number is further tokenized into a sequence of sub-tokens based on the SentencePiece tokenizer \cite{sennrich2016neural}, as did in P5 \cite{p5}. For example, a Yelp item is randomly assigned the number ``4332'', and ``4332'' is represented as a sequence of tokens ``43''``32''.
    \item Title Indexing (TID): Using the item title to represent the item which is also tokenized by SentencePiece \cite{sennrich2016neural}. For example, the Yelp item ``Las Vegas Cigar Outlet'' is represented as a sequence of tokens ``Las''``Vegas''``Ci''``gar''``Outlet''.
    \item Independent Indexing (IID): Creating an independent OOV extra token that needs to be learned for each item. For example, a Yelp item is represented as $\langle$IID5$\rangle$ which is an independent extra token specifically allocated for this item. In the rest of the paper, tokens created for IID will always start with the letters ``IID''.
\end{itemize}

RID generates random numerical indices, leading to potential overlaps between unrelated items after tokenization. For example, two items ``4332'' and ``4389'' would be tokenized into ``43''``32'' and ``43''``89'', respectively, which means that they always share the same sub-token ``43'' even though the two items may be totally unrelated with each other. This unintended overlap may establish arbitrary relationships among items, introducing unwanted bias to model training. As the overlaps stem from the index structure, they are impossible to eliminate no matter how the model learns from data. Consequently, RID is considered an unfavorable method.

TID makes the task more challenging since the model needs to memorize and generate lengthy item titles. Besides, certain words or expressions in the title could be unrelated to the real content of the item, also, very different items may share overlapping tokens in their title, and thus semantics derived from the titles may introduce strong linguistic biases \cite{haowang}. For example, the movies ``The Lord of the Rings'' and ``The Lord of War'' share many tokens in their titles (``the'', ``lord'', ``of''), but they are two very different movies: the former is an epic fantasy, while the later is a crime drama. In general, two irrelevant items could
have very similar titles, such as Apple the fruit and Apple the company, while two closely related items may have very different titles, such as the classic ``beer and diaper'' example in data mining \cite{li2023large}. As a result, using title as ID may encode misleading semantics into the generation process, similar as the problem of random indexing.

IID uses single-token indices for items without assuming any prior information about the items, making the item representations easier for language models to learn compared to RID and TID. Though better than RID and TID, it still has limited performance due to considering all items independent from each other when creating item IDs. It could also incur prohibitively long training time if a large number of new tokens are required to create.

The aforementioned analysis implies that none of the three methods is optimal. To validate this, we provide experimental results to show their suboptimal performance. We evaluate the three indexing methods against two strong and widely-used baselines: SASRec \cite{sasrec} and S$^3$-Rec \cite{zhou2020s3}. Results are shown in Table \ref{tab:basic}, where the best result for each metric is highlighted in bold and the second-best result is underlined with wavy lines. Based on Table \ref{tab:basic}, RID and TID underperform relative to the baselines, while IID offers minor gains at the cost of introducing more learnable tokens because each item is considered as an independent new token. As a result, these indexing methods are considered suboptimal and we will further explore nontrivial indexing methods in the next section.

\section{Nontrivial Indexing Methods}
Based on the above analysis, an optimal item indexing method should meet two criteria to enable an effective learning process:
\begin{enumerate}[leftmargin=*]
    \item Maintaining a suitable length to mitigate the text generation difficulty.
    \item Integrating prior information to item index structure to ensure that similar items share a maximum number of tokens while distinguishable, and dissimilar items share minimal tokens.
\end{enumerate}

To achieve these objectives, we introduce and explore four indexing methods of increasing complexity: Sequential Indexing (SID), Collaborative Indexing (CID), Semantic (content-based) Indexing (SemID), and Hybrid Indexing (HID). SID and CID leverage collaborative information, enabling co-occurring items to share tokens. SemID employs metadata in natural language, allowing semantically similar items to share tokens. HID combines multiple indexing methods, seeking to capitalize on the strengths of each approach in order to generate optimal indices. In the following subsections, we will provide details of the four indexing methods.

\begin{table*}[t]
    \centering
    \begin{tabular}{lcccccccccccc}
    \toprule
    \multicolumn{11}{c}{\textbf{Training Sequence}} & \textbf{Validation} & \textbf{Testing}\\
    \hline
    User 1 & 1001 & 1002 & 1003 & 1004 & 1005 & 1006 & 1007 & 1008 & 1009 & ~ & 1018 & 1019\\ \cline{4-4}  \cline{6-7}
    User 2 & 1010 & 1011 & \multicolumn{1}{|c}{1001} & \multicolumn{1}{|c}{1012} & \multicolumn{1}{|c}{1008} & 1009 & \multicolumn{1}{|c}{1013} & 1014 & ~ & ~ & 1022 & 1023\\\cline{4-7} \cline{10-10}
    User 3 & 1015 & 1016 & 1017 & \multicolumn{1}{|c}{1007} & \multicolumn{1}{|c}{1018} & 1019 & 1020 & 1021 & \multicolumn{1}{|c}{1009} & \multicolumn{1}{|c}{~} & 1015 & 1016\\ \cline{4-6} \cline{10-10}
    User 4 & 1022 & 1023 & \multicolumn{1}{|c}{1005} & 1002 & 1006 & \multicolumn{1}{|c}{1024} & ~ & ~ & ~ & ~ & 1002 & 1008\\ \cline{4-6} \cline{8-10}
    User 5 & 1025 & 1026 & 1027 & 1028 & 1029 & 1030 & \multicolumn{1}{|c}{1024} & {1020} & {1021} & \multicolumn{1}{|c}{1031} & 1033 & 1034\\ \cline{8-10}
    \bottomrule
    \end{tabular}
    \caption{An illustration of Sequential Indexing method. Numbers in boxes represent previously indexed items.}
    \label{tab:sequential_graph}
    \vspace{-20pt}
\end{table*}

\begin{figure*}[t]
  \centering
  \begin{subfigure}[b]{0.5\textwidth}
    \includegraphics[width=\textwidth]{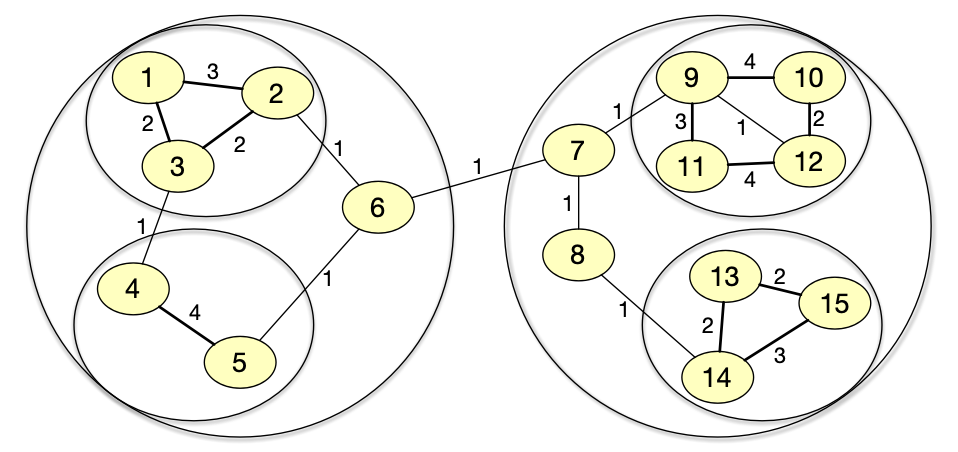}
    \caption{Recursive spectral clustering on item co-appearance graph}
    \label{fig:spectral_clustering}
  \end{subfigure}
  \hfill
  \centering
  \hspace{-20pt}
  \begin{subfigure}[b]{0.22\textwidth}
    \includegraphics[width=\textwidth]{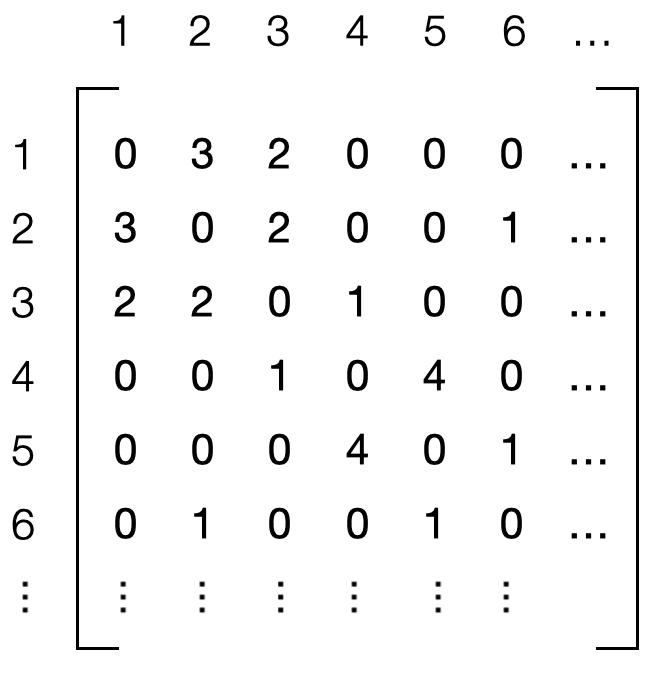}
    \caption{Adjacency matrix}
    \label{fig:adjacency}
  \end{subfigure}
  \hfill
  \centering
  \hspace{-20pt}
  \begin{subfigure}[b]{0.22\textwidth}
    \includegraphics[width=\textwidth]{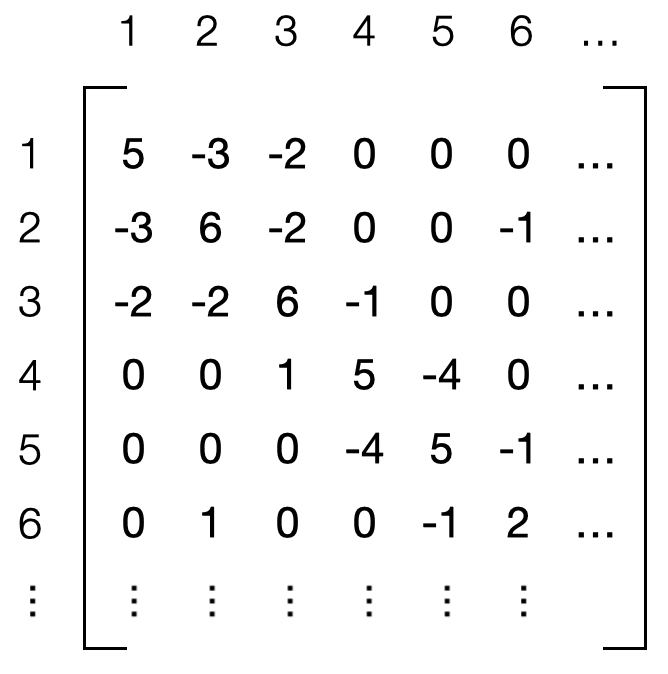}
    \caption{Laplacian matrix}
    \label{fig:laplacian}
  \end{subfigure}
  \vspace{-5pt}
  \caption{Illustration of spectral clustering on the item co-appearance graph based on spectral matrix factorization}
  \label{fig:CID}
  \vspace{-10pt}
\end{figure*}

\subsection{Sequential Indexing}
\label{sec:sequential_indexing}
Sequential indexing is a straightforward method to leverage collaborative information for item indexing. Items interacted consecutively by a user are assigned consecutive numerical indices, reflecting their co-occurrence. Take Table \ref{tab:sequential_graph} as an example, items are assigned with IDs consecutively starting from the first user and all the way to the last user. If an item has already been indexed in previous users' interaction sequence, such as item 1001 in User 2's sequence (and all other squared items in the table), then the item's already assigned ID will be used, otherwise, an incremental new ID will be created and assigned to the item. Notice that the item indexing process only depends on the training sequences, while the validate and testing items do not participate in the indexing process. After the indexing process is finished, the validation and testing items are assigned the corresponding IDs that have already been established during the indexing process.
Upon tokenization based on the SentencePiece tokenizer \cite{sennrich2016neural}, an ID such as ``1001'' will be tokenized into ``100''``1'', while ``1002'' will be tokenized into ``100''``2'', resulting in the shared token ``100'' for these two consecutive items. This gives us encoding similarity between those items that co-appear in at least one user's sequence. As a result, this simple sequential indexing method is able to capture collaborative information on some occasions.

One minor note is that we initiate item index enumeration at 1001. We initiate at 1001 instead of 1 for two reasons: 1) the SentencePiece tokenizer does not tokenize some numbers smaller than 1000 into multiple sub-tokens, such as the number 12, and thus items assigned with these small numbers will be completely independent of each other, 2) after tokenization, smaller numbers could become complete subsets of larger tokenized numbers, e.g., ID ``12'' can be a subset of ID ``12''``34'', which may enforce false correlation between items.

Nevertheless, sequential indexing also has limitations: 1) Adjacently indexed items not interacted together by the same user may erroneously share tokens; for instance, the last item of User 2 is indexed as 1014 (tokenized as ``10''``14'') and the first item of User 3 is indexed as 1015 (tokenized as ``10''``15''), then the token ``10'' will be shared despite a lack of co-occurrence between the two items, 2) it cannot capture similarities based on the frequency of co-occurrence; for example, suppose items 1001 and 1002 co-occur once while items 1002 and 1003 co-occur ten times, both pairs will still share only one token, failing to convey frequency information, and 3) user ordering in the training data affects the results; for example, if we exchange the rows of User 1 and User 2 in Table \ref{tab:sequential_graph}, then the indexing result would be different. Although sequential indexing has its shortcomings, it can still yield relatively favorable results that are close to, or even surpass, the baselines.

\subsection{Collaborative Indexing}
Sequential Indexing is a preliminary method for integrating collaborative information into item indexing. To effectively capture the essence of collaborative filtering, we explore the Collaborative Indexing (CID) approach, which employs spectral clustering based on Spectral Matrix Factorization (SMF) \cite{von2007tutorial,ng2001spectral} to generate item indices. This method is based on the premise that items with more frequent co-occurrence are more similar and should share more overlapping tokens in index construction. The core concept involves constructing a co-occurrence graph for all items based on the training dataset and using spectral clustering to group items into clusters, ensuring that items within the same cluster share tokens when constructing indices.

\subsubsection{Spectral Clustering based on Spectral Matrix Factorization}

To elaborate, we create a graph based on the training set, as exampled in Figure \ref{fig:CID}(a): each item serves as a node, edges between two items represent their co-occurrence (i.e., two items co-appear in a user's interaction sequence), and the edge weights indicate the frequency of co-occurrence (i.e., the number of user interaction sequences in which two items co-appear). The adjacency matrix corresponding to the graph (Figure \ref{fig:CID}(b)) indicates the similarity between items in terms of co-appearance frequency, and the Laplacian matrix corresponding to the graph (Figure \ref{fig:CID}(c)) can be factorized to enable spectral clustering \cite{von2007tutorial,ng2001spectral}.
The spectral clustering process groups items into clusters so that items sharing more co-appearance similarity are grouped into the same cluster; each cluster can be further grouped into finer-grained clusters by recursively applying the spectral clustering process within the big cluster, resulting in hierarchical levels of clusters, as shown in Figure \ref{fig:CID}(a).

More specifically, spectral clustering leverages the eigenvectors of the Laplacian matrix to group nodes into clusters \cite{von2007tutorial,ng2001spectral}. 
It ensures that items within the same cluster have a higher degree of similarity while items in different clusters exhibit lower similarity. We use the standard spectral clustering implementation in the Python scikit-learn package\footnote{\url{https://scikit-learn.org/stable/modules/generated/sklearn.cluster.SpectralClustering.html}}. We do not expand too many details of the spectral clustering algorithm since it is considered a textbook-level algorithm for data analysis \cite{leskovec2020mining}. However, we do want to discuss the two important parameters that are used to control the recursive clustering process: 
1) $N$: we divide the items into $N$ clusters at each level of the clustering, and 2) $k$: the maximum number of items allowed in the final cluster, which serves as  the stopping criterion of the recursive clustering process, i.e., when a cluster contains at most $k$ items, we will not further reduce its size.

Finally, the clustering result can be formulated into a hierarchical tree structure, as shown in Figure \ref{fig:CF_index}. In this figure, each non-leaf node (large yellow nodes in the graph) represents the clusters created at the corresponding level, and each leaf node (small blue nodes) represents an item in the corresponding final cluster. In the next subsection, we will introduce how to create item IDs based on the hierarchical tree structure.

\subsubsection{Item Indexing based on the Spectral Clustering Tree}

\begin{figure}[t]
    \hspace{-15pt}
    \includegraphics*[scale=0.5]{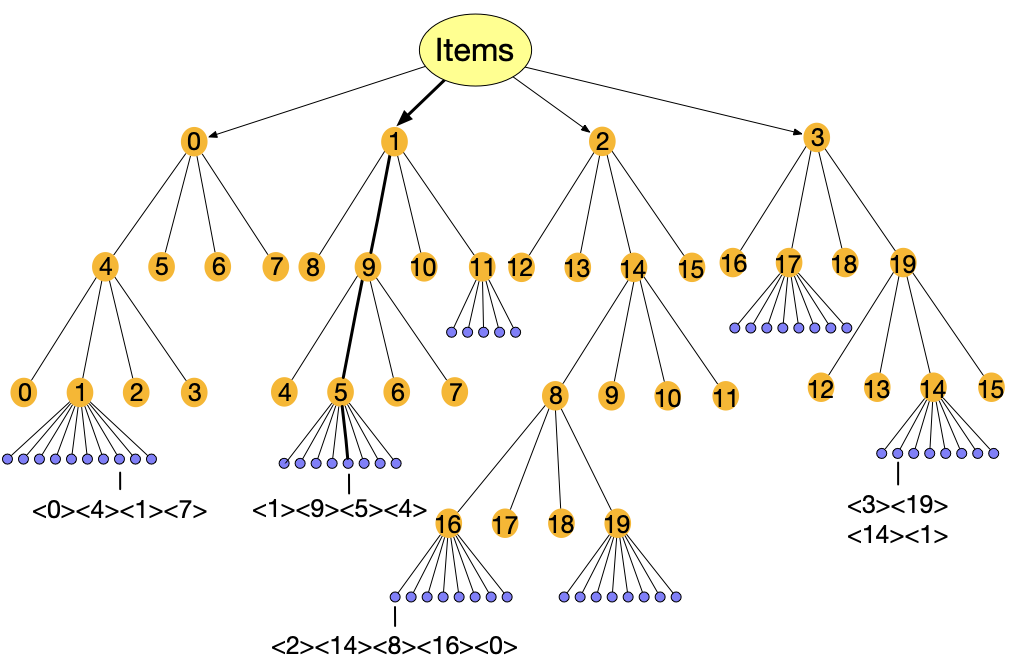}
    \vspace{-15pt}
    \caption{Collaborative indexing based on the spectral clustering tree ($N=4$, $k=20$).}
    \label{fig:CF_index}
    \vspace{-10pt}
\end{figure}

As mentioned above, the recursive clustering process generates a tree structure for the clusters and items, as shown in Figure \ref{fig:CF_index} using $N = 4$ and $k = 20$ as an example,
which means that each iteration of spectral clustering divides items into 4 clusters, and the process is recursively applied on each cluster until the cluster size is smaller than or equal to 20.  
Each non-leaf node (large yellow node) represents a cluster while all items present as leaf nodes (small blue nodes) under the final cluster. Note that since the maximum number of items allowed in the final cluster is $k$, it means that we only need at most $k$ independent extra tokens to distinguish the items within the same final cluster (i.e., the small blue nodes under the same yellow node is at most $k$). As a result, we introduce $k$ independent extra tokens into the vocabulary, noted as $\langle0\rangle$, $\langle1\rangle$, $\langle2\rangle$, $\cdots$, $\langle k-1\rangle$.

We first assign tokens to the non-leaf nodes. The non-leaf nodes are enumerated level by level across the whole tree using the $k$ independent tokens beginning from $\langle0\rangle$ to $\langle k-1\rangle$, as shown in Figure \ref{fig:CF_index}. Once all $k$ tokens are used, we simply restart from $\langle0\rangle$. As mentioned before, each parent cluster node has $N$ children cluster nodes. However, if $N>k$, then we would not have enough tokens to distinguish the different children under the same parent node. As a result, we require $N \leq k$ for collaborative indexing. Together with the level-by-level token assignment process, this can guarantee that different children nodes under the same parent node are assigned different tokens.

We then assign tokens to leaf nodes (small blue nodes), where each leaf node is an item. This is rather straightforward: for each final cluster, we assign each of its children item node with an independent extra token, beginning from $\langle0\rangle$ and on-wards. Since the clustering process ensures that each final cluster contains at most $k$ items, so the $k$ independent extra tokens are enough to distinguish different items under the same final cluster.

Finally, the ID of an item is the concatenation of its non-leaf ancestor nodes' tokens and its own leaf node token. For example, the item under the bolded path in Figure \ref{fig:CF_index} is indexed as $\langle1\rangle\langle9\rangle\langle5\rangle\langle4\rangle$.
This indexing process guarantees that any two items within the same final cluster will share tokens until their own token within the final cluster, which means that the more frequently two items co-occur, the more tokens they will share, well leveraging the collaborative information hidden in user behavior sequences.

\subsection{Semantic (Content-based) Indexing}
\begin{figure}[t]
    \centering
    \includegraphics*[scale=0.5]{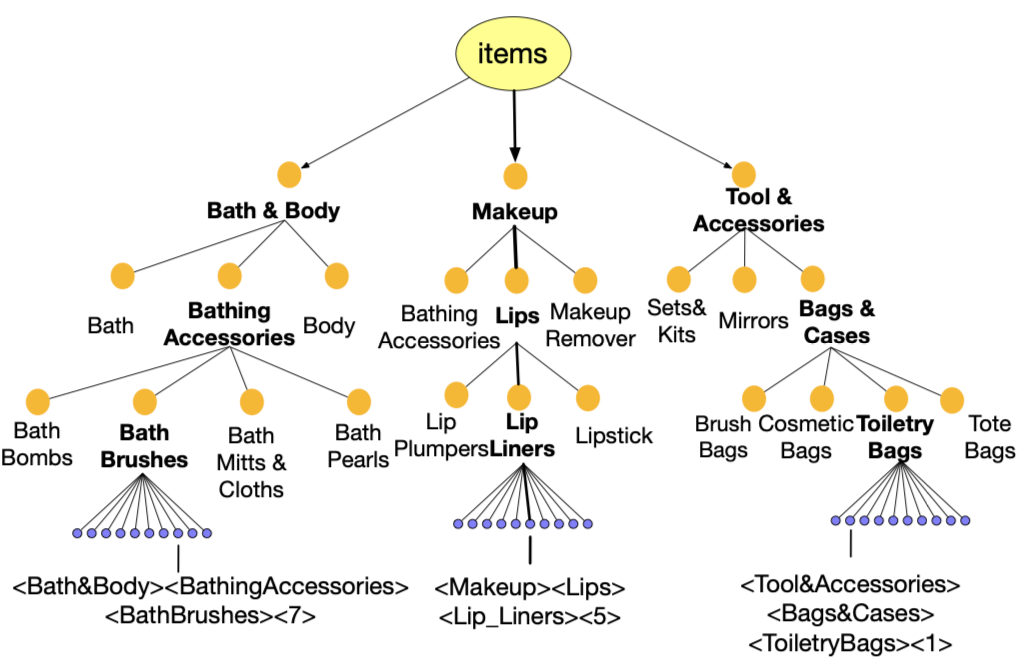}
    \vspace{-20pt}
    \caption{An example of semantic indexing}
    \label{fig:sid_example}
    \vspace{-15pt}
\end{figure}

Semantic (content-based) Indexing (SemID) utilizes item metadata to construct IDs for items.
As shown in Figure \ref{fig:sid_example}, items' categories form a hierarchical structure \cite{zhu2018learning}, with each non-leaf node (large yellow node) representing a category and each leaf node (small blue node) representing an item. Each non-leaf node is assigned an independent extra token, and each leaf node receives a unique extra token under its parent node. To create an item index, the tokens of non-leaf nodes and leaf nodes are concatenated along the path from root to leaf. Take the bolded path in Figure \ref{fig:sid_example} as an example, the item's categories range from coarse to fine-grained as $\langle$Makeup$\rangle$, $\langle$Lips$\rangle$, $\langle$Lip\_Liners$\rangle$, and its leaf node token is $\langle$5$\rangle$, which differentiates this item from other items under the Lip Liners category, then the item would be indexed as $\langle$Makeup$\rangle$$\langle$Lips$\rangle$$\langle$Lip\_Liners$\rangle$$\langle$5$\rangle$.

\subsection{Hybrid Indexing}

Hybrid Indexing (HID) is not a single specific indexing method but a category of methods. It concatenates multiple indices introduced above into one index,
such as SID+IID, CID+IID, SemID+IID, SemID+CID, etc. This approach aims to leverage the advantages of different indexing techniques to produce better indices. In this paper we implement four combinations and here are the details:

For \textbf{SID+IID}: we append an independent extra token at the end of the sequential ID for each item. Suppose the SID of an item after tokenization is ``10''``18'', and its IID index is $\langle$IID982$\rangle$, then the HID index will be ``10''``18''$\langle$IID982$\rangle$. Thus it contains some item co-appearance information from SID and meanwhile ensure the item distinction through IID.

For \textbf{CID} and \textbf{SemID}, before we concatenate them with IID, we first remove the last token (the leaf node token) from them since the last token simply functions to differentiate an item from others under the same parent non-leaf node.
For \textbf{CID+IID}: suppose an item's CID is $\langle1\rangle\langle9\rangle\langle5\rangle\langle4\rangle$, and its IID is $\langle$IID28$\rangle$, then the item's HID would be $\langle1\rangle\langle9\rangle\langle5\rangle\langle$IID28$\rangle$.
For \textbf{SemID+IID}: suppose an item's SemID is $\langle$Makeup$\rangle$$\langle$Lips$\rangle$$\langle$Lip\_Liners$\rangle$$\langle$5$\rangle$, and its IID is $\langle$IID1023$\rangle$, then the HID is $\langle$Makeup$\rangle$$\langle$Lips$\rangle$$\langle$Lip\_Liners$\rangle$$\langle$IID1023$\rangle$. The final index incorporates both collaborative information from CID (or metadata content information in SID), and a special IID token that differentiates the item from all others, ensuring item distinction while preserving the advantages of the CID (or SID). 

For \textbf{SemID+CID}: we concatenate the SemID and CID in either order, hoping to combine both metadata content information and collaborative information. Since both SemID and CID contain leaf node tokens to distinguish items under one parent node, we only need to retain one of them, e.g., we retain the CID leaf node token. Suppose the SemID is $\langle$Makeup$\rangle$$\langle$Lips$\rangle$$\langle$Lip\_Liners$\rangle$$\langle$5$\rangle$ and the CID is $\langle1\rangle\langle9\rangle\langle5\rangle\langle4\rangle$. If we put SemID first, the final HID index is $\langle$Makeup$\rangle$$\langle$Lips$\rangle$$\langle$Lip\_Liners$\rangle$$\langle1\rangle\langle9\rangle\langle5\rangle\langle4\rangle$; otherwise, the HID index is $\langle1\rangle\langle9\rangle\langle5\rangle\langle4\rangle\langle$Makeup$\rangle$$\langle$Lips$\rangle$$\langle$Lip\_Liners$\rangle$.

In the following experiments, we will evaluate and compare the various different HIDs.

\section{Experiments}

\subsection{Dataset and Baselines}
The datasets and their pre-processing methods have been introduced in Section \ref{sec:dataset}. In this section, we introduce the baselines. We apply the various item indexing methods into the P5 framework \cite{p5} for sequential recommendation and compare with several representative sequential recommendation methods as baselines:
\textbf{Caser} \cite{tang2018personalized}: This approach treats sequential recommendation as a Markov Chain and utilizes convolutional neural network to model user interests. \textbf{HGN} \cite{ma2019hierarchical}: This approach leverages hierarchical gating networks to learn user behaviors from both long-term and short-term perspectives. \textbf{GRU4Rec} \cite{jannach2017recurrent}: Originally proposed for session-based recommendation, this approach employs GRU to model the user click history sequence. \textbf{BERT4Rec} \cite{bert4rec}: This approach mimics BERT-style masked language modeling, learning a bidirectional representation for sequential recommendation. \textbf{FDSA} \cite{zhang2019feature}: Focusing on feature transition patterns, this approach models the feature sequence with a self-attention module. \textbf{SASRec} \cite{sasrec}: Adopting a self-attention mechanism in a sequential recommendation model, this approach reconciles the properties of Markov Chains and RNN-based approaches. \textbf{S$^3$-Rec} \cite{zhou2020s3}: Leveraging self-supervised objectives on meta information of items, this approach helps the sequential recommendation model to better discover the correlations among different items and their attributes.
For comparison, we utilize the implementation of S$^3$-Rec and its baselines.

\begin{table*}[t]
    \centering
    \setlength{\tabcolsep}{2pt}
    \begin{tabular}{lcccccccccccc}
    \toprule
        \multirow{2}{*}{Method} & \multicolumn{4}{c}{\textbf{Amazon Sports}} & \multicolumn{4}{c}{\textbf{Amazon Beauty}} & \multicolumn{4}{c}{\textbf{Yelp}}\\
        \cmidrule(lr){2-5} \cmidrule(lr){6-9} \cmidrule(lr){10-13}
         & HR@5 & NCDG@5 & HR@10 & NCDG@10 & HR@5 & NCDG@5 & HR@10 & NCDG@10 & HR@5 & NCDG@5 & HR@10 & NCDG@10 \\
        \midrule
        Caser & 0.0116 & 0.0072 & 0.0194 & 0.0097 & 0.0205 & 0.0131 & 0.0347 & 0.0176 & 0.015 & 0.0099 & 0.0263 & 0.0134 \\
        HGN & 0.0189 & 0.0120 & 0.0313 & 0.0159 & 0.0325 & 0.0206 & 0.0512 & 0.0266 & 0.0186 & 0.0115 & 0.0326 & 0.159 \\
        GRU4Rec & 0.0129 & 0.0086 & 0.0204 & 0.0110 & 0.0164 & 0.0099 & 0.0283 & 0.0137 & 0.0176 & 0.0110 & 0.0285 & 0.0145 \\
        BERT4Rec & 0.0115 & 0.0075 & 0.0191 & 0.0099 & 0.0203 & 0.0124 & 0.0347 & 0.0170 & 0.0051 & 0.0033 & 0.0090 & 0.0090 \\
        FDSA & 0.0182 & 0.0122 & 0.0288 & 0.0156 & 0.0267 & 0.0163 & 0.0407 & 0.0208 & 0.0158 & 0.0098 & 0.0276 & 0.0136 \\
        SASRec & 0.0233 & 0.0154 & 0.0350 & 0.0192 & 0.0387 & 0.0249 & 0.0605 & 0.0318 & 0.0170 & 0.0110 & 0.0284 & 0.0147 \\
        S$^3$\text{-Rec} & 0.0251 & 0.0161 & 0.0385 & 0.0204 & 0.0387 & 0.0244 & 0.0647 & 0.0327 & 0.0201 & 0.0123 & 0.0341 & 0.0168 \\
        \midrule
        RID & 0.0208 & 0.0122 & 0.0288 & 0.0153 & 0.0213 & 0.0178 & 0.0479 & 0.0277  & \underline{0.0225} & \underline{0.0159} & 0.0329 & \underline{0.0193}\\
        TID & 0.000 & 0.000 & 0.000 & 0.000 & 0.0182 & 0.0132 & 0.0432 & 0.0254 & 0.0058 & 0.0040 & 0.0086 & 0.0049\\
        IID & \underline{0.0268} & 0.0151 & \underline{0.0386} & 0.0195 & \underline{0.0394} & \underline{0.0268} & 0.0615 & \underline{0.0341} & \underline{0.0232} & \underline{0.0146} & \underline{0.0393} & \underline{0.0197}\\
        \midrule
        SID & \underline{0.0264} & \underline{0.0186} & 0.0358 & \underline{0.0216} & \underline{0.0430} & \underline{0.0288} & 0.0602 & \underline{0.0368} & \textbf{0.0346} & \textbf{0.0242} & \textbf{0.0486} & \textbf{0.0287}\\
        CID & \uwave{0.0313} & \uwave{0.0224} & \uwave{0.0431} & \uwave{0.0262} & \underline{0.0489} & \underline{0.0318} & \underline{0.0680} & \underline{0.0357} & \underline{0.0261} & \underline{0.0171} & \underline{0.0428} & \underline{0.0225}\\
        SemID & \underline{0.0274} & \underline{0.0193} & \underline{0.0406} & \underline{0.0235} & \underline{0.0433} & \underline{0.0299} & \underline{0.0652} & \underline{0.0370} & \underline{0.0202} & \underline{0.0131} & 0.0324 & \underline{0.0170}\\
        \midrule
        SID+IID & 0.0235 & 0.0161 & 0.0339 & 0.0195 & \underline{0.0420} & \underline{0.0297} & 0.0603 & \underline{0.0355} & \uwave{0.0329} & \uwave{0.0236} & \underline{0.0465} & \uwave{0.0280}\\
        CID+IID & \textbf{0.0321} & \textbf{0.0227} & \textbf{0.0456} & \textbf{0.0270} & \textbf{0.0512} & \textbf{0.0356} & \textbf{0.0732} & \textbf{0.0427} & \underline{0.0287} & \underline{0.0195} & \uwave{0.0468} & \underline{0.0254}\\
        SemID+IID & \underline{0.0291} & \underline{0.0196} & \underline{0.0436} & \underline{0.0242} & \uwave{0.0501} & \uwave{0.0344} & \uwave{0.0724} & \uwave{0.0411} & \underline{0.0229} & \underline{0.0150} & \underline{0.0382} & \underline{0.0199}\\
        SemID+CID & 0.0043 & 0.0031 & 0.0070 & 0.0039 & 0.0355 & 0.0248 & 0.0545 & 0.0310 & 0.0021 & 0.0016 & 0.0056 & 0.0029\\
        \bottomrule
    \end{tabular}
    \caption{Performance of all baseline results and all indexing methods under P5. Numbers in bold represent the best results, numbers with a wavy underline represent the second-best results, and numbers with a straight underline indicate that they are better than the best baseline result. Results better than baselines here have been tested to be significant under the paired Student's t-test protocol with $p$-value $<0.05$.}
    \label{tab:main}
    \vspace{-20pt}
\end{table*}

\subsection{Implementation Details}
Following the P5 framework \cite{p5}, our implementation utilizes T5 as the backbone 
\cite{raffel2020exploring}: there are 6 layers for both encoder and decoder, the model dimensionality is 512 with 8-headed attention.
For tokenization, we use the default SentencePiece tokenizer \cite{sennrich2016neural} with a vocabulary size of 32,128 for parsing sub-word units. All independent extra tokens are not further tokenized.
We use the same sequential recommendation prompts as P5 \cite{p5} to convert sequential information into texts.
We pre-train P5 for 20 epochs using AdamW optimizer on two NVIDIA RTX A5000 GPUs with a batch size of 64, a peak learning rate of 1e-3. We apply warm-up for the first 5$\%$ of all training steps to adjust the learning rate.

RID, TID, and SID do not involve creating OOV tokens since their item indices comprise tokens from the default T5 tokenizer, while IID, CID, SemID, and HID involve creating extra OOV tokens, extending the original vocabulary. All tokens used in these indexing methods, excluding TID, are randomly initialized rather than using T5's pre-trained embeddings for initialization. This is due to our observation that the pre-trained T5's a priori semantics about numbers adversely impact the learning of item semantics and the recommendation performance during experimentation. We use T5's pre-trained token embeddings for initializing TID tokens since TID only involves plain word tokens.

\subsection{Overall Results}
The overall experimental results are presented in Table \ref{tab:main} with all baselines. The best result for each metric is highlighted in bold, while the second-best result is underlined with wavy lines. For each indexing method, if the result surpasses the best baseline result, it is emphasized by underlining with straight lines. In general, RID, TID and IID cannot beat the baseline results in most cases, while most of the advanced indexing methods (SID, CID, SemID and the HIDs) surpass the baseline results. A more detailed breakdown analysis is as follows. 

In Table \ref{tab:main}, the first block contains all the baseline results. The second block contains the basic indexing methods, where RID and TID consistently perform worse than baselines, while IID in general performs better. The third block contains three advanced indexing methods. We can see that SID performs worse than CID and SemID on Amazon datasets but better on Yelp, while CID performs better than SemID across different datasets, indicating that constructing indices using collaborative information is more beneficial than using metadata, because CID can better capture item relationships from user behaviors by collaborative learning from the wisdom of the crowd, which could be more effective than only using items' metadata. The fourth block in the table contains HID results with several different implementations: SID+IID, CID+IID, SemID+IID, and SemID+CID. CID+IID and SemID+IID perform much better than all other indexing methods while SID+IID and SemID+CID perform worse. In the following subsections, we will further analyze the results in the third and fourth blocks in detail based on more comprehensive experiments.

\begin{table*}[t]
    \centering
    \setlength{\tabcolsep}{2pt}
    \begin{tabular}{lcccccccccccc}
    \toprule
        \multirow{2}{*}{Method} & \multicolumn{4}{c}{\textbf{Amazon Sports}} & \multicolumn{4}{c}{\textbf{Amazon Beauty}} & \multicolumn{4}{c}{\textbf{Yelp}}\\
        \cmidrule(lr){2-5} \cmidrule(lr){6-9} \cmidrule(lr){10-13}
         & HR@5 & NCDG@5 & HR@10 & NCDG@10 & HR@5 & NCDG@5 & HR@10 & NCDG@10 & HR@5 & NCDG@5 & HR@10 & NCDG@10 \\
        \midrule
        SASRec & 0.0233 & 0.0154 & 0.0350 & 0.0192 & 0.0387 & 0.0249 & 0.0605 & 0.0318 & 0.0170 & 0.0110 & 0.0284 & 0.0147 \\
        S$^3$\text{-Rec} & 0.0251 & 0.0161 & \uwave{0.0385} & 0.0204 & 0.0387 & 0.0244 & \textbf{0.0647} & 0.0327 & 0.0201 & 0.0123 & 0.0341 & 0.0168 \\
        \midrule
        SID-TSO & \uwave{0.0264} & \uwave{0.0186} & 0.0358 & \uwave{0.0216} & \textbf{0.0430} & \textbf{0.0288} & \uwave{0.0602} & \textbf{0.0368} & \textbf{0.0346} & \textbf{0.0242} & \textbf{0.0486} & \textbf{0.0287}\\
        SID-RO & 0.0214 & 0.0150 & 0.0291 & 0.0175 & 0.0392 & 0.0257 & 0.0512 & 0.0335 & 0.0324 & 0.0219 & 0.0461 & 0.0263\\
        SID-S2LO & \textbf{0.0304} & \textbf{0.0230} & \textbf{0.0395} & \textbf{0.0259} & 0.0395 & 0.0259 & 0.0520 & 0.0337 & \uwave{0.0335} & \uwave{0.0237} & 0.0442 & \uwave{0.0277}\\
        SID-L2SO & 0.0244 & 0.0176 & 0.0356 & 0.0209 & \uwave{0.0409} & \uwave{0.0286} & 0.0586 & \uwave{0.0343} & 0.0316 & 0.0215 & \uwave{0.0472} & 0.0265 \\
        \bottomrule
    \end{tabular}
    \caption{Different settings of Sequential Indexing for P5 compared with two baselines on three datasets. The numbers in bold represent the best results, while the numbers with a wave represent the second-best results. TSO results in Amazon Beauty and Yelp are tested to be significant with respect to other settings.}
    \label{tab:sequential}
    \vspace{-15pt}
\end{table*}

\subsection{Different Settings of Sequential Indexing}
Table \ref{tab:main} shows that though simple in nature, SID can generate favorable results that are close to or surpass baselines. In Section \ref{sec:sequential_indexing}, we explored the construction of SID and its limitations, specifically, the indexing result can be influenced by the user ordering, e.g., if we exchange the rows of User 1 and User 2 in Table \ref{tab:sequential_graph}, then the indexing result would be different. 
In this section, we present the results of SID using four different user orderings, which substantiate this claim and also suggest the most effective ordering to use:
\begin{enumerate}[leftmargin=*]
    \item \textbf{Time-Sensitive Ordering (TSO)}: Users are ordered chronologically in the original dataset based on their initial interaction with the system. Subsequent interactions are recorded and new records are created for previously unrecorded users upon their first interaction with the system. By sorting and processing interactions based on their timestamps, we ensure that users with earlier initial interactions are recorded first.
    \item \textbf{Random Ordering (RO)}: Users are ordered randomly.
    \item \textbf{Short-to-Long Ordering (S2LO)}: Users are organized according to their number of interactions, arranged in ascending order from the fewest to the most interactions.
    \item \textbf{Long-to-Short Ordering (L2SO)}: Users are sorted in descending order from the most to the fewest interactions.
\end{enumerate}

\begin{figure}[t]
  \centering
  \hspace{-20pt}
  \includegraphics[scale=0.2]{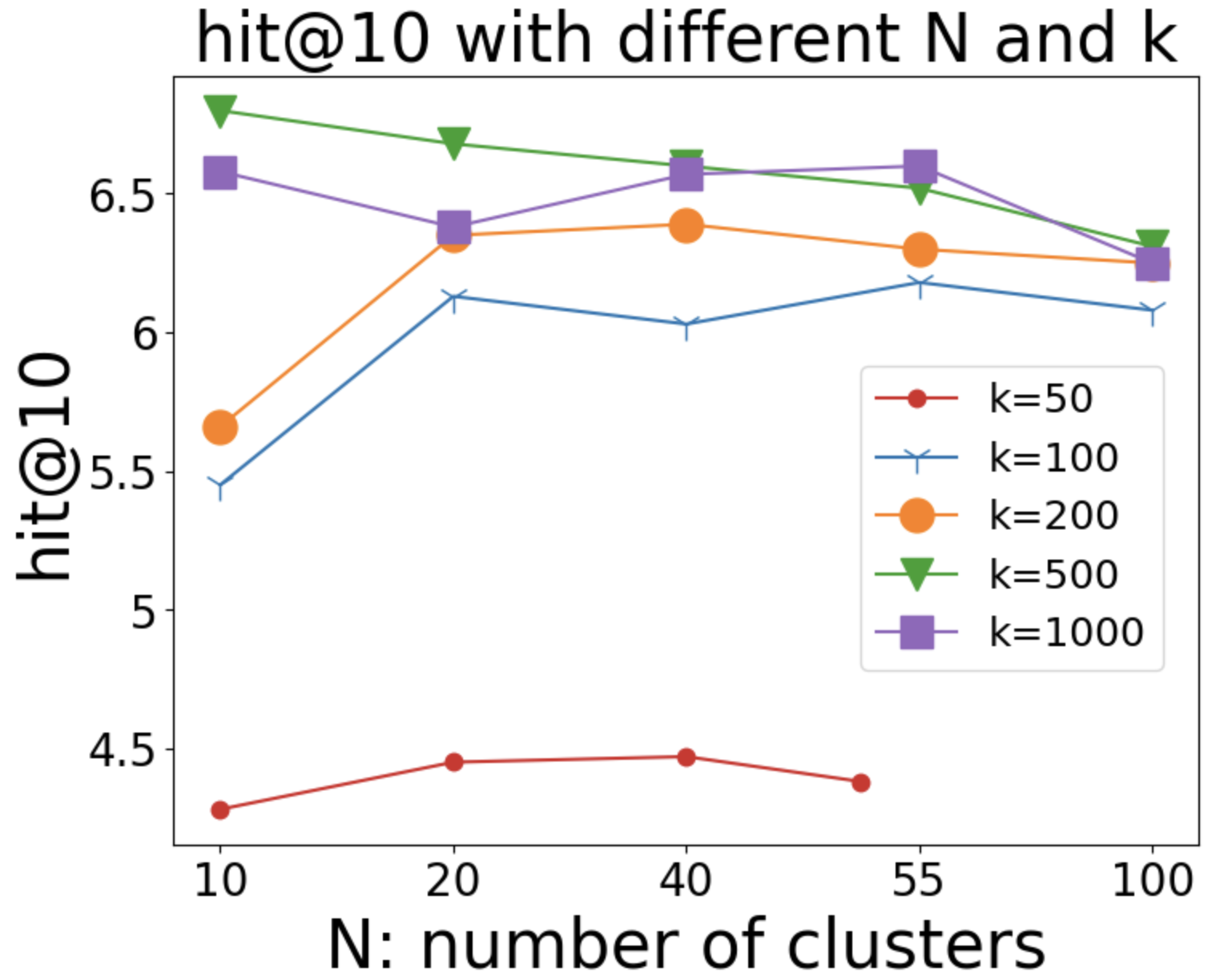}
  \vspace{-5pt}
  \caption{CID Beauty ablations on $N$ (number of clusters at each level) and $k$ (maximum number of items allowed in the final cluster).}
  \label{fig:CID_ablation}
  \vspace{-15pt}
\end{figure}

\begin{figure}[t]
  \centering
  \includegraphics[scale=0.25]{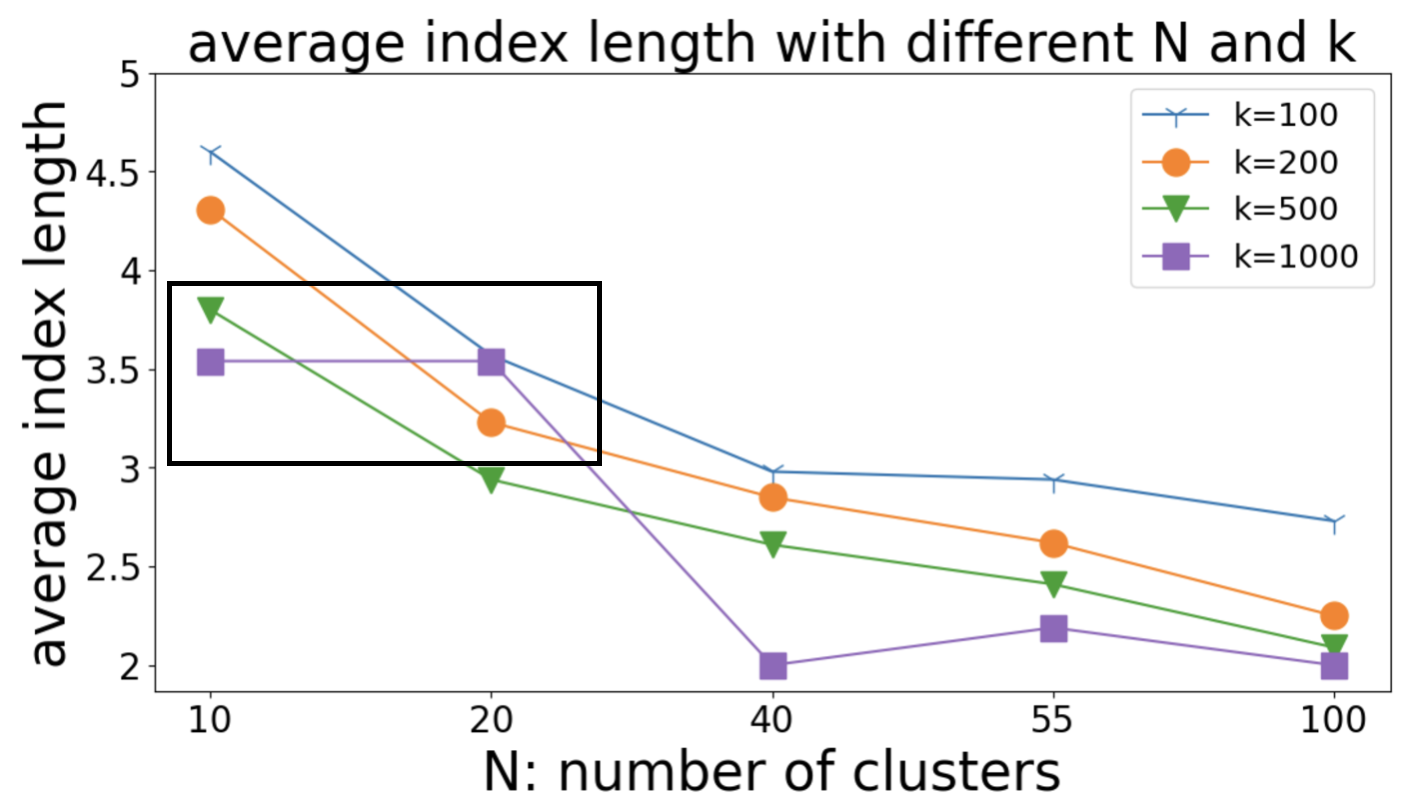}
  \vspace{-5pt}
  \caption{CID average length on Beauty.}
  \label{fig:CID_beauty_length}
  \vspace{-15pt}
\end{figure}

\begin{table}[t] 
\setlength{\tabcolsep}{4pt}
      \begin{tabular}{l|c|c|c|c|c|c}
          \toprule
          \textbf{Dataset} & \multicolumn{2}{c|}{\textbf{Sports}} & \multicolumn{2}{c|}{\textbf{Beauty}} & \multicolumn{2}{c}{\textbf{Yelp}}\\
          \midrule
          SASRec & \multicolumn{2}{c|}{0.0350}& \multicolumn{2}{c|}{0.0605} & \multicolumn{2}{c}{0.0284}  \\
          S$^3$\text{-Rec} & \multicolumn{2}{c|}{0.0385} & \multicolumn{2}{c|}{0.0647} & \multicolumn{2}{c}{0.0341} \\
          \midrule
           & $N$ = 10 & $N$ = 20 & $N$ = 10 & $N$ = 20 & $N$ = 10 & $N$ = 20 \\
           \hline
          $k$=200  & 0.0302 & 0.0423 & 0.0566 & 0.0635 & \uwave{0.0416} & \textbf{0.0428}\\
          $k$=500  & 0.0400 & \uwave{0.0431} & \textbf{0.0680} & \uwave{0.0668}  & 0.0388 & 0.0403\\
          $k$=1000  & \textbf{0.0435} & 0.0416 & 0.0658 & 0.0638 & 0.0385 & 0.0388 \\
          \bottomrule
      \end{tabular}
      \caption{CID hit@10 results under different parameters and datasets. Bold numbers are best results and under-wave numbers are second-best results. The highest scored settings in all datasets are tested to be significant with respect to other settings under the paired Student's t-test with $p$-value < 0.05.} 
      \label{tab:CF}
      \vspace{-20pt}
\end{table} 

\begin{table}[t] 
\setlength{\tabcolsep}{4pt}
      \begin{tabular}{l|c|c|c|c|c|c}
        \toprule
        \textbf{Dataset} & \multicolumn{2}{c|}{\textbf{Sports}} & \multicolumn{2}{c|}{\textbf{Beauty}} & \multicolumn{2}{c}{\textbf{Yelp}}\\
        \midrule
         & $N$ = 10 & $N$ = 20 & $N$ = 10 & $N$ = 20 & $N$ = 10 & $N$ = 20 \\
         \hline
        $k$=200  & 4.25 & 3.35 & 4.31 & 3.23 & \uwave{3.88} & \textbf{3.25} \\
        $k$=500  & 3.66 & \uwave{3.66} & \textbf{3.80} & \uwave{2.94} & 3.57 & 2.91 \\
        $k$=1000  & \textbf{3.31} & 2.78 & 3.54 & 3.54 & 3.21 & 2.76 \\
        \bottomrule
      \end{tabular}
      \caption{Average ID lengths under different parameters. Bold numbers in this table correspond to the best results in Table \ref{tab:CF} (i.e., bold numbers in Table \ref{tab:CF}).} 
      \label{tab:CF_length}
      \vspace{-20pt}
\end{table} 

\begin{table*}[t]
    \centering
    \begin{tabular}{l|l}
    \toprule
        Beauty > Skin Care > \textbf{Eyes} > Combinations & Beauty > Skin Care > Eyes > \textbf{Creams} \\
        Beauty > Makeup > Makeup Remover > \textbf{Eyes} & Beauty > Makeup > Body > Moisturizers > \textbf{Creams} \\
        \bottomrule
    \end{tabular}
    \caption{Examples of non-tree structure categories in Amazon Beauty dataset.}
    \vspace{-15pt}
    \label{tab:nontree}
\end{table*}

\begin{table*}[t]
    \centering
    \setlength{\tabcolsep}{1.5pt}
    \resizebox{\linewidth}{!}{
    \begin{tabular}{lcccccccccccc}
    \toprule
        \multirow{2}{*}{Method} & \multicolumn{4}{c}{\textbf{Amazon Sports}} & \multicolumn{4}{c}{\textbf{Amazon Beauty}} & \multicolumn{4}{c}{\textbf{Yelp}}\\
        \cmidrule(lr){2-5} \cmidrule(lr){6-9} \cmidrule(lr){10-13}
         & HR@5 & NCDG@5 & HR@10 & NCDG@10 & HR@5 & NCDG@5 & HR@10 & NCDG@10 & HR@5 & NCDG@5 & HR@10 & NCDG@10 \\
        \midrule
        SASRec & 0.0233 & 0.0154 & 0.0350 & 0.0192 & 0.0387 & 0.0249 & 0.0605 & 0.0318 & 0.0170 & 0.0110 & 0.0284 & 0.0147 \\
        S$^3$\text{-Rec} & 0.0251 & 0.0161 & 0.0385 & 0.0204 & 0.0387 & 0.0244 & 0.0647 & 0.0327 & \uwave{0.0201} & \uwave{0.0123} & \textbf{0.0341} & \uwave{0.0168} \\
        \midrule
        SemID-non-tree & \textbf{0.0281} & \uwave{0.0192} & \textbf{0.0410} & \uwave{0.0233} & \uwave{0.0423} & \uwave{0.0288} & \uwave{0.0632} & \uwave{0.0354} & 0.0028 & 0.0019 & 0.0050 & 0.0025 \\
        SemID-tree & \uwave{0.0274} & \textbf{0.0193} & \uwave{0.0406} & \textbf{0.0235} & \textbf{0.0433} & \textbf{0.0299} & \textbf{0.0652} & \textbf{0.0370} & \textbf{0.0202} & \textbf{0.0131} & \uwave{0.0324} & \textbf{0.0170} \\
        \bottomrule
    \end{tabular}
    }
    \caption{SemID results under different settings. Bold numbers are best results and under-wave numbers are second-best. Tree setting results in Amazon Beauty and Yelp are tested to be significant with respect to non-tree setting.}
    \label{tab:semantics}
    \vspace{-15pt}
\end{table*}

Table \ref{tab:sequential} presents the performance of the four settings. Our observations indicate that, in general, the relative performance is as follows: Time-Sensitive $>$ $\{$Long-to-Short, Short-to-Long$\}$ $>$ Random. The observations imply that time plays an important role in sequential indexing: items that are interacted at similar times, even by different users, may be more similar to each other compared to items being interacted at vastly different times. As a result, items that occurred at similar times are more likely to be co-interacted by certain users. Thus, using the time-related information when ordering users is likely to improve the performance.

\textbf{Considering these observations, we recommend that future implementations of the simple SID method consider using the time-sensitive user ordering strategies to enhance performance}. Note that the original Amazon and Yelp datasets already used a time-sensitive ordering to arrange the users. As a result, to generate indices using SID, we simply need to incrementally index the items from the first user all the way to the last user.

\subsection{Different Settings of Collaborative Indexing}

CID involves two hyper-parameters: $N$ and $k$, where $N$ is the number of clusters at each level of the clustering, and $k$ is the maximum number of items allowed in the final cluster. Varying these hyper-parameters results in different numbers of independent extra tokens and recommendation performances.

In Figure \ref{fig:CID_ablation}, we present hit@10 results for various $N$ and $k$ value combinations on the Beauty dataset. When $k$ = 50, the performance is below 4.5\%, which is significantly lower than the baselines and some basic indexing methods. However, when $k$ is greater than 100, the performances improve considerably. Furthermore, Table \ref{tab:CF} shows hit@10 results for multiple configurations with $k\in \{200,500,1000\}$, $N\in \{10, 20\}$ and on all three datasets. In these different settings, nearly all the CID results outperform the baselines, indicating that CID is relatively easy to fine-tune with respect to its hyper-parameters.

Based on our observations, we can draw the following conclusions: (1) Extremely small $k$ values lead to suboptimal performance regardless of the chosen $N$. When $k$ = 50, the performance is below the baselines. This can be attributed to the limited expressiveness of a small number of new tokens, which cannot adequately capture the diversity of items. (2) Different $k$ and $N$ combinations yield varying ID lengths (i.e., the number of tokens in an ID). We compute the average ID length for each $k$ and $N$ hyper-parameter setting, and the results are shown in Figure \ref{fig:CID_beauty_length} (for Beauty) and Table \ref{tab:CF_length} (for all datasets). Combining Figure \ref{fig:CID_ablation} and \ref{fig:CID_beauty_length}, as well as Table \ref{tab:CF} and \ref{tab:CF_length}, we find that the optimal recommendation results are usually observed when the average ID length is between 3 and 4. 
For example, the squared points in Figure \ref{fig:CID_beauty_length} shows all cases whose average ID length is between 3 and 4 for the Beauty dataset, and we can see that these points also correspond to the optimal performance on each line in Figure \ref{fig:CID_ablation}. Similarly, the best or second-best results in Table \ref{tab:CF} also corresponds to 3$\sim$4 ID lengths in Table \ref{tab:CF_length} in most cases.

\textbf{Based on these observations, we recommend that future CID implementations use hyperparameters that generate an average ID length between 3 and 4. However, it is worth noting that different datasets may require slightly different lengths for optimal performance.}

\subsection{When will Semantic Indexing Work}
SemID uses metadata to construct item indices. In our experiments, we observe that if the categories follow a hierarchical tree structure, then the performance tends to improve. Category information in datasets is usually not a tree structure because in some cases, one category name can occur under different parent categories, which makes the categories into a graph but not a tree. Table \ref{tab:nontree} are two examples in Amazon Beauty, where the category ``Eyes'' occurs under both ``Skin Care'' and ``Makeup Remover'', and the category ``Creams'' occurs under both ``Skin Care'' and ``Moisturizers''.

To test whether the tree structure in categories is crucial, we compare two different settings in our experiments:
\begin{enumerate}[leftmargin=*]
    \item Non-tree-structure setting: we directly use the category names to create the corresponding independent OOV extra tokens. For example, an item under ``Beauty'', ``Skin Care'', ``Eyes'', and another item under ``Beauty'', ``Makeup'', ``Makeup Remover'', ``Eyes'' will share the token $\langle$Eyes$\rangle$. 
    \item Tree-structure setting: we enforce a tree structure on the categories by creating different OOV tokens when the same category name occurs at different places. For example, the category ``Eyes'' under ``Beauty'', ``Skin Care'' will correspond to token $\langle$Eyes1$\rangle$ while that under ``Beauty'', ``Makeup'', ``Makeup Remover'' corresponds to $\langle$Eyes2$\rangle$.
\end{enumerate}

Table \ref{tab:semantics} illustrates the importance of hierarchical information for SemID's effectiveness. The more closely the categories adhere to a hierarchical structure, the better the performance of the model. This is likely because a hierarchically organized category list helps reduce the search space during the generation process. \textbf{Consequently, this finding highlights the importance of properly organizing and structuring category information when implementing SemID in recommendation foundation models.}

\subsection{What Types of HID Work and Why}
Based on the results presented in Table \ref{tab:main}, CID+IID and SemID+IID show much better performance compared to their respective CID and SemID counterparts. But SID+IID does not improve on SID, and SemID+CID not only does not improve but decreases the performance a lot. Both \textbf{CID+IID} and \textbf{SemID+IID} are constructed by assigning each item an independent extra token and concatenating it after the sequence of cluster IDs or category IDs. These combinations maintain the original index lengths while preserving the hierarchical structure. The improved performance can be attributed to the increased expressiveness of the indices provided by the extra token, as well as the retention of either collaborative information or metadata information within the hybrid index. This combination of factors contributes to the performance enhancement observed in CID+IID and SemID+IID methods.

\textbf{SID+IID} is created by appending an independent extra token after the original sequential index, increasing the ID length by 1. SID+IID does not improve the performance possibly because the additional token interferes the time-sensitive information encoded as a numerical style in the original sequential indices. \textbf{SemID+CID}, which is created by concatenating category IDs with cluster IDs or vice versa, exhibits suboptimal performance, as shown in Table \ref{tab:main}. This holds true for both concatenation orders: category IDs followed by CID indices and cluster IDs followed by SemID indices. The reason behind this suboptimal performance is that it generates excessively long indices and disrupts the hierarchical structure encoded in both SemID and CID. \textbf{Considering our findings, we recommend employing CID+IID and SemID+IID as hybrid indices for recommendation foundation models, as they have demonstrated superior performance in such scenarios.}

\section{Conclusion}
This paper examines various indexing methods using P5 as an example backbone model. We examine three trivial indexing methods: Random Indexing (RID), Title Indexing (TID), and Independent Indexing (IID), and emphasize their limitations. This highlights the importance of selecting an appropriate indexing method for foundation recommendation models, as it greatly impacts the model performance. We then examine four simple yet effective indexing methods: Sequential Indexing (SID), Collaborative Indexing (CID), Semantic Indexing (SemID), and Hybrid Indexing (HID). Experimental results on Amazon Sports, Amazon Beauty, and Yelp datasets demonstrate their strong performance. The four effective indexing methods satisfy the two criteria introduced in this paper: (1) maintaining a suitable ID length, and (2) integrating useful prior information into item ID construction. We hope this study serves as an inspiration for future research on indexing methods for recommendation foundation models and beyond.\\

\noindent\textbf{Acknowledgement:} The work was supported in part by NSF IIS-2046457 and IIS-2007907. 

\bibliographystyle{ACM-Reference-Format}
\bibliography{main}

\end{document}